\documentclass[%
reprint,
nofootinbib,
 amsmath,amssymb,
 aps,
 pra,
]{revtex4-1}

\usepackage{mathrsfs}

\usepackage{graphicx}
\usepackage{dcolumn}
\usepackage{bm}

\usepackage{amsthm}
\usepackage{xfrac}

\usepackage{url}
\usepackage{hyperref}
\usepackage{color}
\definecolor{refcolor}{RGB}{0,0,190}
\hypersetup{
    colorlinks,
    citecolor=refcolor,
    filecolor=refcolor,
    linkcolor=refcolor,
    urlcolor=refcolor
}
\usepackage{bookmark}
\bookmarksetup{
  numbered, 
  open,
}

\newtheorem{problem}{Problem}
\newtheorem{postulateHS}{Postulate}
\renewcommand{\thepostulateHS}{HS\arabic{postulateHS}}
\newtheorem{postulatePS}{Postulate}
\renewcommand{\thepostulatePS}{PS\arabic{postulatePS}}
\makeatletter
\newcommand{\setpostulateHStag}[1]{
  \let\oldthepostulateHS\thepostulateHS
  \renewcommand{\thepostulateHS}{#1}
  \g@addto@macro\endpostulateHS{
    \global\let\thepostulateHS\oldthepostulateHS}
  }
\makeatother
\makeatletter
\newcommand{\setpostulatePStag}[1]{
  \let\oldthepostulatePS\thepostulatePS
  \renewcommand{\thepostulatePS}{#1}
  \g@addto@macro\endpostulatePS{
    \global\let\thepostulatePS\oldthepostulatePS}
  }
\makeatother

\newtheorem{problemO}{Objectivity Problem}

\newtheorem{definition}{Definition}

\newtheorem{criterion}{Criterion}

\theoremstyle{remark}

\begin{document}


\newcommand{\pbref}[1]{\ref{#1} (\nameref*{#1})}
   
\def\({\left(}
\def\){\right)}

\newcommand{\tn}{\textnormal}
\newcommand{\ds}{\displaystyle}
\newcommand{\dsfrac}[2]{\displaystyle{\frac{#1}{#2}}}

\newcommand{\statespace}{\mathcal{S}}
\newcommand{\hilbert}{\mathcal{H}}
\newcommand{\vectorspace}{\mathcal{V}}
\newcommand{\mc}[1]{\mathcal{#1}}
\newcommand{\ms}[1]{\mathscr{#1}}

\newcommand{\wh}[1]{\widehat{#1}}
\newcommand{\wt}[1]{\widetilde{#1}}
\newcommand{\wht}[1]{\widehat{\widetilde{#1}}}
\newcommand{\on}[1]{\operatorname{#1}}

\newcommand{\qmU}{$\mathscr{U}$}
\newcommand{\qmR}{$\mathscr{R}$}
\newcommand{\qmUR}{$\mathscr{UR}$}
\newcommand{\qmDR}{$\mathscr{DR}$}

\newcommand{\R}{\mathbb{R}}
\newcommand{\C}{\mathbb{C}}
\newcommand{\Z}{\mathbb{Z}}
\newcommand{\K}{\mathbb{K}}
\newcommand{\N}{\mathbb{N}}
\newcommand{\Prj}{\mathcal{P}}
\newcommand{\abs}[1]{\left|#1\right|}

\newcommand{\de}{\operatorname{d}}
\newcommand{\tr}{\operatorname{tr}}
\newcommand{\im}{\operatorname{Im}}

\newcommand{\ie}{\textit{i.e.}\ }
\newcommand{\vs}{\textit{vs.}\ }
\newcommand{\eg}{\textit{e.g.}\ }
\newcommand{\cf}{\textit{cf.}\ }
\newcommand{\etc}{\textit{etc}}
\newcommand{\etal}{\textit{et al.}}

\newcommand{\Span}{\tn{span}}
\newcommand{\pde}{PDE}
\newcommand{\U}{\tn{U}}
\newcommand{\SU}{\tn{SU}}
\newcommand{\GL}{\tn{GL}}

\newcommand{\schrod}{Schr\"odinger}
\newcommand{\vonneum}{Liouville-von Neumann}
\newcommand{\ks}{Kochen-Specker}
\newcommand{\leggarg}{Leggett-Garg}
\newcommand{\bra}[1]{\left\langle#1\right|}
\newcommand{\ket}[1]{\left|#1\right\rangle}
\newcommand{\braket}[2]{\langle#1|#2\rangle}
\newcommand{\ketbra}[2]{\left|#1\rangle\langle#2\right|}
\newcommand{\expectation}[1]{\langle#1\rangle}
\newcommand{\Herm}{\tn{Herm}}
\newcommand{\Sym}[1]{\tn{Sym}_{#1}}
\newcommand{\meanvalue}[2]{\langle{#1}\rangle_{#2}}

\newcommand{\btimes}{\boxtimes}
\newcommand{\btimess}{{\boxtimes_s}}

\newcommand{\h}{\mathbf{(2\pi\hbar)}}
\newcommand{\x}{\mathbf{x}}
\newcommand{\z}{\mathbf{z}}
\newcommand{\q}{\mathbf{q}}
\newcommand{\p}{\mathbf{p}}
\newcommand{\0}{\mathbf{0}}
\newcommand{\annih}{\widehat{\mathbf{a}}}

\newcommand{\cs}{\mathscr{C}}
\newcommand{\ps}{\mathscr{P}}
\newcommand{\xhat}{\widehat{\x}}
\newcommand{\phat}{\widehat{\mathbf{p}}}
\newcommand{\fqproj}[1]{\Pi_{#1}}
\newcommand{\cqproj}[1]{\wh{\Pi}_{#1}}
\newcommand{\cproj}[1]{\wh{\Pi}^{\perp}_{#1}}

\newcommand{\M}{\mathbb{E}_3}
\newcommand{\D}{\mathbf{D}}
\newcommand{\dn}{\tn{d}}
\newcommand{\db}{\mathbf{d}}
\newcommand{\n}{\mathbf{n}}
\newcommand{\m}{\mathbf{m}}
\newcommand{\V}[1]{\mathbb{V}_{#1}}
\newcommand{\F}[1]{\mathcal{F}_{#1}}
\newcommand{\Fvacuumfield}{\widetilde{\mathcal{F}}^0}
\newcommand{\nD}[1]{|{#1}|}
\newcommand{\Lin}{\mathcal{L}}
\newcommand{\End}{\tn{End}}
\newcommand{\vbundle}[4]{{#1}\to {#2} \stackrel{\pi_{#3}}{\to} {#4}}
\newcommand{\vbundlex}[1]{\vbundle{V_{#1}}{E_{#1}}{#1}{M_{#1}}}
\newcommand{\rep}{\rho_{\scriptscriptstyle\btimes}}

\newcommand{\intl}[1]{\int\limits_{#1}}

\newcommand{\moyalBracket}[1]{\{\mskip-5mu\{#1\}\mskip-5mu\}}

\newcommand{\Hint}{H_{\tn{int}}}

\newcommand{\quot}[1]{``#1''}

\def\sref #1{\S\ref{#1}}

\newcommand{\dBB}{de Broglie--Bohm}
\newcommand{\dBBt}{{\dBB} theory}
\newcommand{\pwt}{pilot-wave theory}
\newcommand{\PWT}{PWT}
\newcommand{\NRQM}{{\textbf{NRQM}}}

\newcommand{\image}[3]{
\begin{center}
\begin{figure*}[!ht]
\includegraphics[width=#2\textwidth]{#1}
\caption{\small{\label{#1}#3}}
\end{figure*}
\end{center}
}

\title{Standard Quantum Mechanics without observers}

\author{Ovidiu Cristinel Stoica}
\affiliation{
 Dept. of Theoretical Physics, NIPNE---HH, Bucharest, Romania. \\
	Email: \href{mailto:cristi.stoica@theory.nipne.ro}{cristi.stoica@theory.nipne.ro},  \href{mailto:holotronix@gmail.com}{holotronix@gmail.com}
	}%

\date{\today}

\begin{abstract}
The Projection Postulate from Standard Quantum Mechanics relies fundamentally on measurements. But measurements implicitly suggest the existence of anthropocentric notions like measuring devices, which should rather emerge from the theory. This article proposes an alternative formulation of the Standard Quantum Mechanics, in which the Projection Postulate is replaced with a version in which measurements and observations are not assumed as fundamental. More precisely, the Wigner functions representing the quantum states on the phase space are required to be tightly constrained to regions of the classical coarse-graining of the phase space. This ensures that states are quasiclassical at the macro level. Within a coarse-graining region, the time evolution of the Wigner functions representing the quantum system is required to obey the Liouville-von Neumann equation, the phase-space equivalent of the Schr\"odinger equation. The projection is postulated to happen when the system transitions from a coarse-graining region to others, by selecting one of them according to the Born rule, but without reference to a measurements. The connection with the standard formulation of Quantum Mechanics is explained, as well as the problems that the present formulation solves, in particular the Wigner's friend type of paradoxes. Experimental consequences and open problems of the proposed formulation are discussed.
\end{abstract}


\maketitle

\begin{quote}
\emph{I like to think the moon is there even if I am not looking at it.}
\end{quote}
\begin{flushright}
Albert Einstein
\end{flushright}

\section{Introduction}
\label{s:intro}

Standard Quantum Mechanics (SQM) in its various formulations relies essentially on measurements, which assume at least implicitly the existence of quantum measuring devices as classical or at least \emph{quasiclassical} systems, \ie systems that at the macro level appear to be classical.
Here by SQM I mean the familiar textbook QM viewed as a complete, self-contained description of the world.
The central role of measurements is manifest in the \emph{Projection Postulate}, which states that the observed quantum system is found to be in an eigenstate of the operator associated to the observable we measure, and prescribes the probability for each outcome to occur according to the \emph{Born rule}.
For each quantum measurement, one assumes that the world is divided in two parts, one is classical or quasiclassical, and includes the apparatus, and the other one is the observed quantum system. Bohr prescribed that the apparatus is a classical system. John von Neumann treated the apparatus like a quantum system which behaves quasiclassically, but gave a central role to the observer, and the split between the apparatus and the observed system persisted \cite{vonNeumann1955MathFoundationsQM}. This split remains true for modern approaches to quantum measurement \cite{BuschLahtiMittelstaedt1996TheQuantumTheoryOfMeasurement}.
Moreover, since the SQM leads to macro superpositions like {\schrod} cats, something is needed to project large systems to make them appear classical. In the Copenhagen Interpretation, this is achieved by the observer, whose sensory organs or maybe consciousness act like measuring devices.
It seems that the measurement process, and implicitly the observer, play a fundamental role in the theory.
This leads to some foundational problems.

\begin{problem}[of observer]
\label{pb:Observer}
Can the Postulates of Quantum Mechanics be formulated without relying on measurements and/or observations?
\end{problem}

The apparently privileged role of the measurement process was noticed and bothered various researchers. {\schrod} \cite{schrodinger1935SchrodingerCat} and Wigner \cite{Wigner1967RemarksOnTheMindBodyProblemWignersFriend} conceived thought experiments, known as the \emph{{\schrod}'s cat} paradox and \emph{Wigner's friend} paradox, aiming to emphasize the problem. Since measurements imply the existence of measuring devices and observers, a rich literature appeared, aiming to dethrone their apparently central role, and to explain why, in a fundamentally quantum world, what we perceive is apparently classical. An analysis of the problem of the observer and proposed interpretations aiming to remove it from its central position can be found in \cite{Goldstein1998QuantumTheoryWithoutObserversI,Goldstein1998QuantumTheoryWithoutObserversII}. Maudlin even ``banned'' the Copenhagen Interpretation from his book \emph{Philosophy of Physics: Quantum Theory} \cite{Maudlin2019PhilosophyofPhysicsQuantumTheory}, motivating
that ``it is not even in the running for a description of the physical world and what it does.'' Here I will take a rather constructive position and propose a reformulation of SQM aiming to absolve it of this problem, without aiming that no problems remain and without denying the usefulness of more complete descriptions.

Even if a formulation which solves Problem \ref{pb:Observer} can be found, it still needs to be able to describe quantum measurements, in order to connect the theory with the experiments.
Here it is important to introduce the distinction between classical, quantum, and quasiclassical system. A \emph{classical system} is of course a system as described in Classical Mechanics, while a \emph{quantum system} is a system as described in Quantum Mechanics. But since the world is quantum, all systems are quantum, including the measuring devices, and this was never denied. From the beginning, Bohr insistent on treating the measuring devices as classical in relation to using classical concepts and logic in the description of experiments and their outcomes, but he never claimed that they actually are classical systems. Rather, the emphasis was that measuring devices appear and behave classically at the macro level.
In the same spirit, in this article, by \emph{quasiclassical system} I understand a quantum system that at the macro level is indistinguishable from a classical system.
But since a measuring apparatus is a quasiclassical system, another problem appears:
\begin{problem}[of classicality criterion]
\label{pb:ClassicalCriterion}
What conditions should a quantum system satisfy to be considered quasiclassical?
\end{problem}

One of the goals of this article is to address Problem \ref{pb:ClassicalCriterion}. I will give a more precise characterization of quasiclassicality later, in Criterion \ref{criterion-classicality}, but since I will mention it earlier throughout the article, and because it is central to the formulation I will present, let us anticipate it. A (quantum) system is \emph{quasiclassical} if the representation of its quantum state as a Wigner function is sufficiently localized on the phase space to be macroscopically indistinguishable from a point. Since a point in the same phase space represents a classical system, it follows that a quantum system is quasiclassical if it cannot be distinguished from a classical system. Later I will make clearer the connection between the macro level and quasiclassicality, attempting thus to implement, in the most conservative way, Bohr's views on Quantum Mechanics, in a way that includes the measuring device in the same quantum world which contains the observed system.

Another problem is that the space of quantum states is vastly large, and most of the states it includes can be rather seen as superpositions of quasiclassical systems. In particular, the {\schrod} equation predicts that, even if we start with a quasiclassical measuring device, during the measurement it evolves into a superposition. Hence, there is a closely related problem:
\begin{problem}[of the classical level]
\label{pb:ClassicalLevel}
Why does the world appear to be classical at the macro level?
\end{problem}

Since the postulates of SQM in the usual formulations do not address Problems \ref{pb:ClassicalCriterion} and \ref{pb:ClassicalLevel}, relying on measurements in the Projection Postulate is in some sense circular: on the one hand, the Projection Postulate makes the world appear classical, on the other hand, the Projection Postulate assumes measurements, which in their turn require  quasiclassical measuring devices.

\begin{problem}[of circularity]
\label{pb:ProjectionCircularity}
How can we avoid the circularity that quasiclassicality is achieved by measurement, but a measurement requires a quasiclassical measuring apparatus?
\end{problem}

The presence of the Projection Postulate in SQM, without specifying exactly the conditions that make the projection happen, is another problem:
\begin{problem}[of projection conditions]
\label{pb:ProjectionCriterion}
Under what conditions does the projection occur?
\end{problem}

Closely related, and due to the fact that SQM does not specify when exactly between the preparation and the measurement the projection takes place, but also because the quantum state seems to depend on the operator corresponding to the observation we make, is the following problem:
\begin{problem}[of state reality]
\label{pb:StateReality}
Can the quantum state be well defined at all times?
\end{problem}

Probably by removing {\schrod} cats, Wigner's friend paradox \cite{Wigner1967RemarksOnTheMindBodyProblemWignersFriend} is eliminated, but stronger versions of the paradox suggest that this is not enough \cite{FrauchigerRenner2018QMNoConsistentDescription,Brukner2018NoGoObserverIndependentWignerFriend,BongEtAl2020StrongNoGoTheoremWignerFriend}.
\begin{problem}[of Wigner's friend]
\label{pb:WignerFriend}
Can the \emph{Wigner's friend} type of paradoxes be avoided?
\end{problem}

These problems are often considered to be of interest just to the philosophy of physics, in particular to ontology, and addressing them is usually done by the so-called ``interpretations'' of QM, still considered by many working physicists not serious science.
But it is preferable for a theory in physics to be able to describe precisely and unambiguously both its states and its dynamics. And if it can do this without appealing to measurements in its very postulates, much of the discussion about foundations can move into the field of physics, where it belongs, rather than philosophy.

Possible answers to at least some of the problems mentioned above are already proposed by various interpretations of QM, and are extensively developed and researched.
For example, in the \emph{objective collapse theories} quantum states collapse into sufficiently well localized, hence quasiclassical, states \cite{GhirardiRiminiWeber1986GRWInterpretation,sep-qm-collapsetheories}. This is accomplished by modifying the dynamics of QM with randomly occurring collapses. 
We do not know yet if a modification of the dynamics is needed. It is believed that \emph{decoherence} can solve these problems without modifying the dynamics \cite{JoosZehKieferGiuliniKupschStamatescu2003DecoherenceAndTheAppearanceOfClassicalWorldInQuantumTheory,zurek1981pointer,Zurek1982EnvironmentInducedSuperselectionRules,Zurek2003EnvironmentAssistedInvarianceEntanglementAndProbabilitiesInQuantumPhysics,Zurek2003DecoherenceEinselectionAndTheQuantumOriginsOfTheClassical,Zurek2006DecoherenceTransitionFromQuantumToClassicalRevisited,Schlosshauer2007Decoherence}. But since decoherence can at best explain the emergence of the classical world by branching, even if this solution will turn out to work, it requires to be accompanied by an interpretation where branching occurs, like Everett's \cite{Everett1957RelativeStateFormulationOfQuantumMechanics,Everett1973TheTheoryOfTheUniversalWaveFunction}, or Consistent Histories \cite{Griffiths1984ConsistentHistories,Omnes1992ConsistentInterpretationsQuantumMechanics,GellMannHartle1990DecoheringHistories,GellMannHartle1990QuantumMechanicsInTheLightOfQuantumCosmology}, or the de Broglie-Bohm theory \cite{Bohm52,Bohm2004CausalityChanceModernPhysics}. 
Despite the extended work done and the progress made with the decoherence program, it is not clear yet that decoherence alone can fully solve these problems \cite{pessoa1997canDecoherence,Adler2003DecoherenceNotSolvedMeasurementProblem,leggett2002limitsQM}. It is considered that a decoherence-based solution of Problems \ref{pb:ClassicalCriterion} and \ref{pb:ClassicalLevel} requires a preferred basis to emerge, which then can be applied to solve Problem \ref{pb:Observer} and the other problems.

In this article, I propose a formulation of SQM which addresses the above mentioned problems in a very straightforward manner.
In the following I will call it \emph{Objective Standard Quantum Mechanics} (OSQM).
Like SQM, OSQM is still based on the {\schrod} dynamics and a projection postulate, but without invoking observations or measurements, and at the same time it assumes a clear prescription of what is quasiclassical and what is the state of the system at any given time.
OSQM does not necessarily compete with the various interpretations of QM, but it rather slightly strengthens SQM. 
The idea is to replace all observations and measurements with a single quasiprojection, so that the system is always in a macro state, where macro or quasiclassicality is defined as in the Classical Statistical Mechanics formulation of Thermodynamics. The justification for this is presented in Sec. \sref{s:justification}.
In \sref{s:postulatesSQM} I recall the postulates of SQM, in \sref{s:objectivity} I justify and add another postulate to impose objectivity. In \sref{s:summary} I sketch OSQM, as an implementation of objectivity, and I explain the necessity to take into account the coarse graining of the classical phase space into macroscopically indistinguishable states.

Fortunately, along with the Heisenberg and {\schrod} pictures, shown by {\schrod} to be equivalent \cite{Schrodinger1926RelationHeisenbergSchrodinger}, it is possible to formulate Quantum Mechanics on the phase space \cite{Weyl1946ClassicalGroupsInvariantsAndRepresentations,Wiger1959GroupTheoryAppsQM,Moyal1949PhaseSpaceQM,Groenewold1946PhaseSpaceQM}. This extends to the Projection Postulate \cite{Baker1958FormulationofQMPhaseSpace}. There are already original phase space formulations like \cite{Dragoman2005PhaseSpaceFormulationMeasurementProblem} and \cite{deGosson2018EmergenceOfQuantumFromClassical}, but the approach presented here is different, by relying on the coarse graining.
In Sec. \sref{s:postulates} I introduce the Postulates of OSQM.
In Sec. \sref{s:projection_postulate} I show that the standard Projection Postulate can be recovered from OSQM, and I give some examples.
A more detailed explanation of the interplay between dynamics and projections is given in Sec. \sref{s:interplay}.
Experimental predictions, and how they are already confirmed, will be presented in Sec. \sref{s:experiment}.
A discussion of how OSQM solves the above mentioned problems, as well as of possible shortcomings and remaining open problems, takes place in Sec. \sref{s:discussion}.
Sec. \sref{s:discussion} also includes a brief comparison with the Consistent Histories approach, another formulation of SQM which does not rely in its Postulates on measurements.

Problem \ref{pb:ClassicalLevel} is solved, albeit only provisionally, by imposing the criterion of classicality proposed to address Problem \ref{pb:ClassicalCriterion} as a fundamental principle, rather than deducing it from other principles. The main difference from SQM consists in using the criterion of classicality (from the proposed solution of Problem \ref{pb:ClassicalCriterion}) to replace the Projection Postulate with a version which does not require observations or measurements, and applies to the total wave function. This resolves Problem \ref{pb:Observer}, avoids the circularity mentioned in Problem \ref{pb:ProjectionCircularity}, defines the conditions required in Problem \ref{pb:ProjectionCriterion}, allows the quantum state to be defined at all times (Problem \ref{pb:StateReality}), and avoids the Wigner's friend type of paradoxes (Problem \ref{pb:WignerFriend}).

\section{Justification}
\label{s:justification}

In this section I show that, if we start from SQM
by which I mean the familiar to us textbook QM viewed as a complete, self-contained description of the world,
and modify it in a way that makes observations objective, we are led naturally to OSQM.

\subsection{Postulates of Standard Quantum Mechanics}
\label{s:postulatesSQM}

First, let us review the Hilbert space Postulates of SQM.

\begin{postulateHS}[Quantum state]
\label{ppSQMQuantumState}
The state of a quantum system is represented by a ray in a Hilbert space $\hilbert$.
\end{postulateHS}

\begin{postulateHS}[Composite system]
\label{ppSQMComposite}
For composite systems, the Hilbert space of states is the tensor product of the Hilbert spaces associated to the component systems.
\end{postulateHS}

\begin{postulateHS}[Observable]
\label{ppSQMQuantumOperator}
An observable of a system is represented by a Hermitian operator on its Hilbert space $\hilbert$.
\end{postulateHS}

\begin{postulateHS}[Dynamics]
\label{ppSQMQuantumDynamics}
As long as no measurement takes place, the evolution of the quantum state $\ket{\psi(t)}$ of the system is described by the {\schrod} equation \eqref{eq:schrod},
\begin{equation}
\label{eq:schrod}
i\hbar\frac{\de}{\de t}\ket{\psi(t)}=\wh{H}(t)\ket{\psi(t)},
\end{equation}
where $\wh{H}(t)$ is the \emph{Hamiltonian operator}. 
\end{postulateHS}

A fundamental role in the usual formulations of SQM is played by the \emph{Projection Postulate}.
Let $\mc{S}$ be the observed quantum system, and $\mc{M}$ the measuring apparatus, having as Hilbert spaces $\hilbert_{\mc{S}}$ and $\hilbert_{\mc{M}}$. 
Let $\wh{A}$ be a Hermitian operator on $\hilbert_{\mc{S}}$, representing the observable measured by the measuring device $\hilbert_{\mc{M}}$. We assume for simplicity that the spectrum of $\wh{A}$ is non-degenerate, and $N=\dim\hilbert_{\mc{S}}$.
The measuring apparatus $\mc{M}$ is assumed, by definition and construction, to be a system that can be in one of the following states:
\begin{enumerate}
	\item 
	A quasiclassical state $\ket{\text{ready}}_{\mc{M}}$, corresponding to the system $\mc{M}$ being prepared to observe.
	\item 
	One of $N$ quasiclassical states $\ket{\text{outcome}=\lambda_j}_{\mc{M}}$, corresponding to the measuring apparatus $\mc{M}$ indicating that the outcome of the measurement is $\lambda_j$, for $j\in\{1,\ldots,N\}$.
\end{enumerate}
All these states are assumed to be mutually orthogonal.
The measuring device is constructed so that, if the observed system is already in an eigenstate $\ket{j}$ of $\wh{A}$ before the measurement, the {\schrod} evolution of the  composed system is, for all $j\in\{1,\ldots,N\}$,
\begin{equation}
\label{eq:measurement_def}
\ket{\text{ready}}_{\mc{M}}\otimes\ket{j}\mapsto\ket{\text{outcome}=\lambda_j}_{\mc{M}}\otimes\ket{j}.
\end{equation}

Then, {\schrod}'s equation predicts that if the observed system is in the initial state
\begin{equation}
\label{eq:observed_initial_state}
\ket{\psi}=\sum_{j=1}^N c_j \ket{j},
\end{equation}
the composed system evolves into 
\begin{equation}
\label{eq:measurement_superposition}
\ket{\text{ready}}_{\mc{M}}\otimes\ket{\psi}\mapsto\sum_{j=1}^N c_j \ket{\text{outcome}=\lambda_j}_{\mc{M}}\otimes\ket{j}.
\end{equation}

With these settings, the Projection Postulate states:
\begin{postulateHS}[Projection Postulate]
\label{ppSQMQuantumTransition}
The result of the measurement is one of the eigenvalues $\lambda_j$ of $\wh{A}$, and the state of the total system is, after the measurement, projected to the state $\ket{\text{outcome}=\lambda_j}_{\mc{M}}\otimes\ket{j}$.
The probability to obtain the outcome $\lambda_j$ is $\abs{c_j}^2$.
\end{postulateHS}

\subsection{Objectivity}
\label{s:objectivity}

To make the observations objective, we have to make sure that all measurements taking place at the same time are compatible. Let the observed systems be $\mc{S}_i$, the observable operators be $\wh{A}_i$, and the Hilbert space of each of the observed systems be $\hilbert_i$. If any two measurements that take place at the same time are measurements of distinct systems $\mc{S}_i$ and $\mc{S}_j$, then the corresponding operators $\wh{A}_i$ and $\wh{A}_j$ are defined on distinct Hilbert spaces $\hilbert_i$ and $\hilbert_j$. Their corresponding operators $\wh{A}_i\otimes I_{\hilbert_j}$ and $I_{\hilbert_i}\otimes\wh{A}_j$ on the Hilbert space $\hilbert_i\otimes\hilbert_j$ commute, ensuring the compatibility of the operators $\wh{A}_i$ and $\wh{A}_j$.

If the observed systems have common parts, the compatibility of two observations is no longer automatic, so it has to be imposed. Let the Hilbert spaces of the systems $\mc{S}_i$ and $\mc{S}_j$ have the form $\hilbert_i=\hilbert'_i\otimes\hilbert_{i \cap j}$ and $\hilbert_j=\hilbert_{i \cap j}\otimes\hilbert'_j$, where $\hilbert_{i \cap j}$ is the Hilbert space of their common part. Then, the observations are compatible only if the operators $\wh{A}_i\otimes I_{\hilbert'_j}$  and $I_{\hilbert'_i}\otimes\wh{A}_j$ commute.

In a similar way, one can require that all measurements taking place at the same time are compatible, and hence that they can be combined into a larger measurement.

However, the measuring device is also an observed system. Following von Neumann \cite{vonNeumann1955MathFoundationsQM}, we can treat it as a quantum system, but in this case the observer behaves as a measuring apparatus for the measuring device itself, and projects it in a quasiclassical state. 
But if we endow the observer with the capacity to project the measuring device into a quasiclassical state, why not making this the reason for the quasiclassicality of the macro level itself? This is quite vague, since we did not define what an observer is, and what allows the observer project systems, including his or her own brain state and body, in quasiclassical states. But let us accept provisionally the existence of a ultimate class of measuring devices responsible for the functionality of the actual measuring devices.

In terms of observers, let $\mc{S}_i$ be the largest system observed by the observer $\mc{O}_i$, which includes the measuring device and the environment that the observer can see, and the body of the observer as well. Whatever remains outside of $\mc{S}_i$, the rest of the universe, is the complementary system $\mc{S}_i^c$. The total Hilbert space of the universe, $\hilbert$, decomposes as $\hilbert=\hilbert_i\otimes\hilbert_i^c$ (I ignore the order of the Hilbert spaces in the tensor product, for simplicity of notation). The operator corresponding to the observer's act of observing the world has the form $\wh{\mc{A}}_i=\wh{A}_i\otimes I_{\hilbert_i^c}$.

The condition that all observers in the universe make compatible observations is then that for any two observers $\mc{O}_i$ and $\mc{O}_j$, the corresponding operators $\wh{\mc{A}}_i$ and $\wh{\mc{A}}_j$ commute. Therefore, the set of operators $\wh{\mc{A}}_i,i\in\mc{I}$ determine a huge joint measurement of the entire universe, and they decompose the total Hilbert space into a direct sum
\begin{equation}
\label{eq:totality_observables}
\hilbert=\bigoplus_\alpha\hilbert_\alpha,
\end{equation}
where any two distinct Hilbert subspaces $\hilbert_\alpha$ and $\hilbert_\beta$ are orthogonal. 

As a way to remove the measurements and observations from the postulates, it seems reasonable to decouple the decomposition from Eq. \eqref{eq:totality_observables} from the existence of the observers, and rather postulate it directly. This would be more general, and able to allow the existence of a macroscopic objective world even in the absence of observers. Usually the postulates do not mention the macro level and the fact that it is quasiclassical, and neither what ``macro'' or ``quasiclassical'' mean. This does not follow from the other postulates. We will come back to this in \sref{s:postulate-macrolevel}, but for the moment let us note that it makes sense that the macro level of reality is a decomposition as in Eq. \eqref{eq:totality_observables}.

However, several problems should be taken into account before postulating a global decomposition like \eqref{eq:totality_observables}.

The first Objectivity Problem is posed by the \emph{quantum Zeno effect} \cite{DegasperisFondaGhirardi1974ZenoEffect,MisraAndSudarshan1977QuantumZenoEffect,Khalfin1958DecayOfAQuasiStationaryStateZenoEffect,WilkinsonEtAl1997NonExponentialDecayInQuantumTunnelingZenoEffect}.
Consider that a succession of measurements of the same observable $\wh{A}$ are performed on a quantum system, at the times $t_1,\ldots,t_n\ldots$. Supposed that after the $n$-th measurement we find the system in the state $\ket{\psi(t_n)}=\ket{j}$, where $\ket{j}$ is an eigenstate of the operator $\wh{A}$. After the measurement, the system evolves into a superposition $\ket{\psi(t)}=\sum_k a_k(t)\ket{k}$ of more eigenstates of $\wh{A}$. Under certain conditions, if the time interval between the measurements is short enough, the probability that the observed system collapses back into the eigenstate $\ket{j}$ is overwhelmingly larger than collapsing into another eigenstate. Reducing the time interval makes the probability as closer to the unity as desired, since the probability to collapse on another eigenstate is proportional to $\Delta t^2$. Thus, by repeated measurement, the observed system can be made to remain indefinitely ``frozen'' in the state $\ket{\psi}=\ket{j}$, justifying the name ``quantum Zeno effect''.
This effect goes back to von Neumann \cite{vonNeumann1955MathFoundationsQM}, who observed that it is possible to steer the evolution of a quantum system by continuous measurements, and to Turing, who noticed that in particular this can be used to freeze the state of the quantum system \cite{Teuscher2004AlanTuringLifeAndLegacy}, therefore it is also called \emph{Turing's paradox}.

\begin{problemO}
\label{pbo:zeno}
If we assume a continuous measurement whose eigenspaces are given by Eq. \eqref{eq:totality_observables}, how can we avoid the quantum Zeno effect, which seems to freeze the dynamics?
While the dimension of a Hilbert space $\hilbert_\alpha$ may in general be infinite, the continuous global measurement seems to confine the system in the same state subspace indefinitely, preventing any dynamics observable at the macro level.
\end{problemO}

Another problem is posed by the \emph{Wigner-Araki-Yanase (WAY) theorem}, which shows that conservation laws prevent in many cases the measurements from being sharp.
Consider the pre-measurement stage, when the observed system and the measuring device evolved as in Eq. \eqref{eq:measurement_superposition}, but the Projection Postulate was not invoked yet. The evolution of the combined system up to this stage should be unitary, hence it should conserve the expectation values of all operators commuting with the Hamiltonian of the combined system. Wigner used this argument to show that this can prevent, in certain situations, the state vectors in superposition in Eq. \eqref{eq:measurement_superposition} from being orthogonal \cite{wigner1952MessungQMOperatoren,Wigner1952MessungQMOperatorenPBusch2010EnTranslation}. 
When this happens, the decomposition \eqref{eq:totality_observables} cannot be exact, and the measurement cannot be sharp. The result was extended by Araki and Yanase \cite{ArakiYanase1960MeasurementofQMOperators}.
\begin{problemO}
\label{pbo:way-theorem}
The decomposition \eqref{eq:totality_observables} assumes sharp measurements, but these are prevented in many cases by the WAY theorem. 
\end{problemO}
\begin{problemO}
\label{pbo:leggett-garg-theorem}
Can a decomposition like \eqref{eq:totality_observables} be consistent with the \emph{Legget-Garg theorem} \cite{LeggettGarg1985QMvsMacrorealism,Leggett2008Realism,EmaryNeillFranco2013LeggettGarg}, which forbids the simultaneous conditions of macrorealism and noninvasive measurements?
\end{problemO}
\begin{problemO}
\label{pbo:macrolevel}
Can there exist something like the decomposition \eqref{eq:totality_observables}, which allows only total quantum states that look classical at the macro level?
\end{problemO}

Objectivity Problem \ref{pbo:zeno} suggests that instead of a decomposition as a direct sum of mutually orthogonal Hilbert spaces as in Eq. \eqref{eq:totality_observables}, one should consider something more general, which allows the state to evolve from one of the terms of the decomposition to another.
Objectivity Problems \ref{pbo:way-theorem} and \ref{pbo:leggett-garg-theorem} suggest the same, that one should not impose a strict decomposition of the Hilbert space as a direct sum. This means that we should find something else instead of a global standard measurement -- also called \emph{projection-valued measures} (PVM).
The natural choice is \emph{generalized measurements}, which use a \emph{positive operator-valued measure} (POVM). This solution allows the replacement of the total operator that gives the decomposition \eqref{eq:totality_observables}, with an over-complete set of positive semidefinite operators, that sum up to the identity operator and define the POVM. If the positive semidefinite operators defining the POVM overlap, this breaks the confinement to Hilbert subspaces, and provides a way to avoid the Objectivity Problems \ref{pbo:zeno}--\ref{pbo:leggett-garg-theorem}.
The question remains how to define the POVM on the total Hilbert space $\hilbert$, so that the macro level looks classical, solving by this the Objectivity Problem \ref{pbo:macrolevel}. This is one of the central points of OSQM, and will be addressed in \sref{s:postulate-macrolevel}. For the moment, let us ignore quasiclassicality, and formulate this constraint as an additional Postulate on top of the Postulates \ref{ppSQMQuantumState}--\ref{ppSQMQuantumTransition} of SQM.

\begin{postulateHS}[Macro objectivity]
\label{ppSQMMacroObjectivity}
There is a POVM on the total Hilbert space $\hilbert$, such that
\begin{enumerate}
	\item 
Its positive semidefinite operators quasiproject any state in $\hilbert$ into a state that looks classical at the macro level.
	\item 
Any ordinary quantum measurement is compatible with the POVM.
\end{enumerate}
\end{postulateHS}

While Postulate \ref{ppSQMMacroObjectivity} may seem complicated, we obtained it naturally, by imposing to SQM the additional requirement of macro objectivity. The problem is to make it precise and to define and implement quasiclassicality. We will see in \sref{s:postulate-macrolevel} that there is a natural way to do this in a way that also provides a natural solution to Objectivity Problem \ref{pbo:macrolevel}.

\subsection{Summary of the proposed formulation OSQM}
\label{s:summary}

I now sketch the way I will apply the phase-space representation of QM to obtain OSQM to address these problems. The Postulates are proposed and discussed in Sec. 
\sref{s:postulates}. We start with a brief review of the classical phase space (in \sref{s:classical-phase-space}) and quantization (in \sref{s:quantization}), to fix the notations. Then, in \sref{s:postulate-quantum-state} and \sref{s:postulate-quantum-observables}, we reformulate the first three Postulates \ref{ppSQMQuantumState},  \ref{ppSQMComposite}, and \ref{ppSQMQuantumOperator} in terms of the phase space. A quantum state is thus described by its Wigner function on the classical phase space. An observable is represented by its Weyl symbol, a real-valued function on the classical phase space.

As a quasiclassical level, I will assume the macro level from Classical Mechanics, which is the \emph{coarse graining} of the classical phase space used by Boltzmann to formulate the Second Law of Thermodynamics in terms of Statistical Mechanics. This coarse graining is the partition of the phase space in equivalence classes of states that are macroscopically indistinguishable. The exact procedural definition of coarse graining is still not well understood, but it works perfectly fine to ``reduce'' Thermodynamics to Statistical Mechanics \cite{Eddington1928NatureOfThePhysicalWorld,Feynman1965TheCharacterOfPhysicalLaw,Penrose1989EmperorsNewMind,schulman1997timeArrowsAndQuantumMeasurement,DavidZAlbert2003TimeAndChance,BarryLoewer2016MentaculusVision}. It makes sense to use it in Quantum Mechanics as well, since (1) the macro state is the same as the classical macro state, and (2) quantum measurements work by ``amplifying''  differences at the quantum level into differences at the quasiclassical macro level.
Therefore, I will replace Postulate \ref{ppSQMMacroObjectivity} with a phase space Postulate discussed in \sref{s:postulate-macrolevel}, which establishes that in the quantum world, the macro states are quasiclassical, in the sense that they are quasirestricted to a coarse-graining region of the classical phase space.
I will clarify exactly what ``quasirestricted'' means. The reason why I am using it, rather than using a ``strict restriction'', will be explained, but we already have an idea from the discussion in \sref{s:objectivity}.

The dynamics within a coarse-graining region is just the {\schrod} dynamics as in Postulate \ref{ppSQMQuantumDynamics}, but expressed on the phase space by the {\vonneum} equation (in \sref{s:postulate-dynamics}). This applies only as long as the system stays within the same coarse-graining region. When it evolves to other coarse-graining regions, the Wigner function is forced to choose one of the coarse-graining regions, which is tantamount to a projection that does not rely on measurements or observations (in \sref{s:postulate-transitions}).
The condition of quasirestriction becomes important here, because unlike strict restrictions, it allows small overlaps of the coarse-graining regions near their boundaries, so that the state can transition from one region to another despite Objectivity Problem \ref{pbo:zeno} from \sref{s:objectivity}.

OSQM is translated back into the more popular Hilbert space formulation in \sref{s:hilbert-space}.

\section{Postulates}
\label{s:postulates}

\subsection{Classical phase space}
\label{s:classical-phase-space}

We review the phase space of a classical system of $\m$ particles in the three-dimensional space $\R^3$.
The classical \emph{configuration space} is the $\n$-dimensional space $\R^\n$, where $\n=3\m$.
The evolution depends not only on the configuration $\x\in\R^\n$, but also of the momenta $p_j=m_j v_j=m_j \dot{x}_j=m_j \dsfrac{\de x_j}{\de t}$. The \emph{state} (or \emph{phase}) of a classical system characterized by both positions and momenta, and it is represented by a point $\z=(\x,\p)$ in the \emph{phase space} $\R^{2\n}=\R^{6\m}$.

The phase space $\R^{2\n}$ is naturally endowed with a \emph{symplectic structure}
\begin{equation}
\label{eq:symplectic_struct}
J=\begin{pmatrix}
O_\n & I_\n \\
-I_\n & O_\n \\
\end{pmatrix},
\end{equation}
which satisfies $J^2=-I$. It defines the \emph{symplectic product} between $\z=(\x,\p)$ and $\z'=(\x',\p')$,
\begin{equation}
\label{eq:symplectic-prod}
\sigma(\z,\z')=\z^T J \z'=\x\cdot\p'-\p\cdot\x'.
\end{equation}

A classical observable is a real function on the phase space, $A:\R^{2\n}\to\R$. One defines the \emph{Poisson bracket} between two classical observables $A,B$ by
\begin{equation}
\label{eq:poisson-bracket}
\{A,B\}(\x,\p):=(\partial_\x A \partial_\p B - \partial_\x B \partial_\p A)(\x,\p).
\end{equation}

The dynamics follows \emph{Hamilton's equations},
\begin{equation}
\label{eq:hamilton}
\begin{cases}
\dot{\x}(t)=\ \ \partial_{\p}H(\x,\p,t)\\
\dot{\p}(t)=-\partial_{\x}H(\x,\p,t),\\
\end{cases}
\end{equation}
where the \emph{Hamilton function} is a scalar function defined on $\R^{2\n}\times\R$.

\subsection{Quantization}
\label{s:quantization}

Quantization associates to each classical observable a Hermitian operator acting on the Hilbert space $\hilbert=L^2(\R^\n,\C)$ consisting of the \emph{square-integrable} complex functions on the configuration space $\cs=\R^\n$.
The Hilbert space $\hilbert$ is endowed with the Hermitian scalar product
\begin{equation}
\label{eq:hermitian-scalar-product}
\braket{\psi}{\phi} := \int_{\R^\n}\overline{\psi(\x)}\phi(\x)\de\x.
\end{equation}

To the constant observable $A(\z)=1$, the quantization procedure associates the identity operator $\wh{I}$, to the position $x_j$ and to the momentum component $p^j$ it associates the operators $\wh{x}_j\ket{\psi}=x_j\ket{\psi}$ and $\wh{p}_j\ket{\psi}=-i\hbar\partial_j\ket{\psi}$. Since the classical observables $x_j$ and $p_j$ commute, and their corresponding quantum operators do not commute, we also need to specify a rule to choose a particular ordering of the products of such observables. In the following we will assume the most commonly used \emph{Weyl quantization rule}, which uses the symmetrized product
\begin{equation}
\label{eq:quantization-rule-weyl}
(x_j)^r(p_j)^s\mapsto\((\wh{x}_j)^r(\wh{p}_j)^s\)_{\tn{sym}}.
\end{equation}

The quantum states are represented by rays in $\hilbert$, but in case they are mixed or the information is incomplete, they can be represented more generally as density operators, which are self-adjoint operators $\rho$ on $\hilbert$, whose diagonal elements are non-negative and add up to $1$.

\subsection{Quantum states}
\label{s:postulate-quantum-state}

Let $\wh{\rho}$ be a \emph{density operator} on $\hilbert$, representing the state of the quantum system. If the system is in a pure state, then $\wh{\rho}=\ket{\psi}\bra{\psi}$, where $\ket{\psi}\in\hilbert$ is the unit vector representing the state of the system.

The \emph{Wigner phase-space function} of $\wh{\rho}$, which will be called in the following the \emph{Wigner function}, is defined as
\begin{equation}
\label{eq:wigner-function}
W_\rho(\x,\p) := \frac{1}{\h^\n}\int_{\R^\n} e^{-\frac{i}{\hbar} \x'\cdot\p}\bra{\x+\frac{\x'}{2}}\wh{\rho}\ket{\x-\frac{\x'}{2}}\de \x'.
\end{equation}

The density operator can be obtained from its Wigner function by
\begin{equation}
\label{eq:density-op-from-wf}
\wh{\rho} = \iiint\ket{\x+\frac{\x'}{2}} e^{\frac{i}{\hbar}\x'\cdot\p} W_\rho(\x,\p) \bra{\x-\frac{\x'}{2}}\de\x \de \p \de\x'.
\end{equation}

In particular, for a pure state $\wh{\rho}=\ket{\psi}\bra{\psi}$,
\begin{equation}
\begin{aligned}
\label{eq:wigner-function-psi}
W_\psi(\x,\p) = & \frac{1}{\h^\n}\int_{\R^\n} e^{-\frac{i}{\hbar} \x'\cdot\p} \\
&\psi\(\x+\frac{\x'}{2}\)\psi^\ast\(\x-\frac{\x'}{2}\)\de \x'.
\end{aligned}
\end{equation}

The \emph{recovery property} holds: if $\psi(\0)\neq 0$, the state $\ket{\psi}$ can be recovered from the Wigner function,
\begin{equation}
\label{eq:psi-from-wigner-function}
\psi(\x)\psi^\ast(\mathbf{0}) = \int_{\R^\n} W_{\psi}\(\frac{\x}{2},\p\) e^{\frac{i}{\hbar} \x \cdot\p}\de \p.
\end{equation}

The Wigner function is real-valued, but can take negative values, and for this reason it cannot be a probability distribution, but it can be a quasiprobability distribution, from which the correct probability distributions in the position and momentum bases can be recovered as marginal distributions.

We introduce the following postulates:
\setpostulatePStag{PS1}
\begin{postulatePS}[Quantum state]
\label{ppPSQuantumState}
The state of a quantum system is represented by a time dependent Wigner function on the classical phase space.
\end{postulatePS}

\setpostulatePStag{PS2}
\begin{postulatePS}[Composite system]
\label{ppPSComposite}
For composite systems, the phase space is the Cartesian product of the phase spaces associated with the component systems.
\end{postulatePS}

\subsection{Observables}
\label{s:postulate-quantum-observables}

Let $\mc{S}(\R^{\n})$ be the space of smooth complex functions $f$ so that, for any multiindices $\alpha,\beta$, $\x^\alpha\partial_\x^\beta f$ is bounded in $\R^n$.
Under the Weyl quantization rule \eqref{eq:quantization-rule-weyl}, to any complex function $A\in \mc{S}(\R^{2\n})$ we associate the operator $\wh{A}$ defined by
\begin{equation}
\label{eq:weyl-operator}
\bra{\x}\wh{A}\ket{\x'} = \frac{1}{\h^\n} \int_{\R^{\n}} e^{\frac{i}{\hbar}(\x-\x')\cdot\p} A\(\frac{\x+\x'}{2},\p\)\de\p.
\end{equation}

The operator $\wh{A}$ acts on a quantum state $\ket{\psi}$ by
\begin{equation}
\label{eq:weyl-operator-act}
\begin{aligned}
\wh{A}\psi(\x) = &\int_{\R^\n} \bra{\x}\wh{A}\ket{\x'}\psi(\x') \de\x'\\
= &\frac{1}{\h^\n}\iint_{\R^{2\n}} e^{\frac{i}{\hbar}(\x-\x')\cdot\p} \\
& A\(\frac{\x+\x'}{2},\p\)\psi(\x')\de\x'\de\p.
\end{aligned}
\end{equation}

The Weyl symbol $A(\x,\p)$ can be obtained from $\wh{A}$:
\begin{equation}
\label{eq:weyl-symbol}
A(\x,\p) = \int_{\R^{\n}} e^{-\frac{i}{\hbar}\x'\cdot\p} \bra{\x+\frac{\x'}{2}}\wh{A}\ket{\x-\frac{\x'}{2}}\de\x'.
\end{equation}

The Weyl correspondence $A \leftrightarrow \wh{A}$ is a linear bijection. We call $A$ the \emph{Weyl symbol} of the operator $\wh{A}$.
Since $\wh{A}^\dagger=\wh{A^\ast}$, $A$ is real iff $\wh{A}$ is Hermitian.

We introduce the following postulate:
\setpostulatePStag{PS3}
\begin{postulatePS}[Observable]
\label{ppPSQuantumOperator}
An observable of the system is represented by a real-valued function on the classical phase space (its Weyl symbol).
\end{postulatePS}

The Weyl symbol of a density operator $\wh{\rho}$ is $\h^\n W_\rho(\x,\p)$.
Hence, the Weyl symbol $\pi_{\ket{\psi}}$ of the projector $\wh{\pi}_{\ket{\psi}}:=\ket{\psi}\bra{\psi}$ is
\begin{equation}
\label{eq:proj}
\pi_{\ket{\psi}}(\x,\p) = \h^\n W_\psi(\x,\p).
\end{equation}

The mean value of an operator $\wh{A}$ is given by
\begin{equation}
\label{eq:mean-value}
\langle\wh{A}\rangle_{\ket{\psi}} = \int_{\R^{2\n}} A(\z)W_{\psi}(\z)\de\z.
\end{equation}

In particular, by applying \eqref{eq:mean-value} to $\wh{\pi}_{\ket{\psi'}}=\ket{\psi'}\bra{\psi'}$,
\begin{equation}
\label{eq:transition}
\abs{\braket{\psi}{\psi'}}^2=\h^\n\int_{\R^{2\n}} W_{\psi}(\z)W_{\psi'}(\z)\de\z.
\end{equation}

The \emph{Moyal product} of two observables  $A,B\in\mc{S}\(\R^{2\n}\)$ is the Weyl symbol of the product $\wh{A}\wh{B}$, and it is given by
\begin{equation}
\begin{aligned}
\label{eq:Moyal-prod-int}
(A\star B)(\z) = & \(\pi\hbar\)^{-2\n}\iint_{\R^{4\n}} e^{-\frac{i}{2\hbar}\sigma\(\z',\z''\)} \\
& A\(\z+\z'\)B\(\z+\z''\)\de\z'\de\z''.
\end{aligned}
\end{equation}
The local form (``local'' on the phase space) of the Moyal product operator is
\begin{equation}
\label{eq:Moyal-prod-local}
\star = e^{\frac{i\hbar}{2}\sigma(\overleftarrow{\partial}_{\z},\overrightarrow{\partial}_{\z})}.
\end{equation}

\subsection{The quasiclassicality of the macro world}
\label{s:postulate-macrolevel}

In order to define what is understood for a quantum state to be quasiclassical, we will rely on the idea that the macro world, even if we know it to be quantum, looks like the classical macro world.
To formalize this idea, we will use the quasiprojection operators proposed by Omn{\`e}s \cite{Omnes1997QuantumClassicalCorrespondenceUsingProjectionOperators}.
We will also take into account the insights of de Gosson regarding the phase space and quantum blobs \cite{deGosson2013QuantumBlobs}, based on the \emph{principle of symplectic camel} discovered by Gromov \cite{Gromov1985SymplecticCamel}.

To define the classical macro level, we assume that the classical theory to which we applied the quantization procedure has a definite set of observables $\mc{M}$, which will be called \emph{macro observables}. They are in general aggregate functions of other observables: averages, integrals or sums, volumes, densities \etc, and are important for example in Statistical Mechanics. The macro observables partition the phase space into \emph{coarse-graining regions} where the macro observables take constant (or indistinguishable) values. Each of these regions of the phase space contains (micro) states that cannot be distinguished macroscopically. Even for a classical theory there are ambiguities in defining exactly the coarse graining, but given the success of Statistical Mechanics, in particular in reducing Thermodynamics to Classical Mechanics, we will assume that both the macro observables and the coarse graining can be defined unambiguously at least in principle.

To be able to transfer the classical coarse graining of the phase space to the quantized theory, it is necessary to impose certain restrictions on the coarse-graining regions. We cannot allow just any subsets of the phase space to represent quasiclassical macro states, because the Wigner function cannot be arbitrarily localized, it has at least to satisfy Heisenberg's uncertainty principle. The projection of the support of the Wigner functions on any plane defined by any pair of conjugate variables $(q_j,p_j)$ should be at least $\dsfrac\hbar2$. Therefore, any partition of the phase space consistent with Quantum Mechanics should consist of unions of \emph{quantum blobs}, introduced by de Gosson \cite{deGosson2013QuantumBlobs}, because
\begin{enumerate}
	\item 
a quantum blob is the smallest symplectic invariant region of the phase space compatible with the uncertainty principle, and 
	\item 
we want to exclude regions containing only functions that are not Wigner functions corresponding to quantum states, being too localized or having negative quasiprobability. 
\end{enumerate}
Quantum blobs have size comparable with the Planck constant, and are in a one-to-one correspondence with the \emph{squeezed coherent states} from SQM.
As such, defining the partitions of the classical phase space as unions of quantum blobs does not imply significant departures from the classical macro states, but they are essential to support quantum states.
Quantum blobs were also used in the formulations or interpretations of quantum mechanics, \eg in \cite{deGosson2018EmergenceOfQuantumFromClassical} and \cite{Dragoman2005PhaseSpaceFormulationMeasurementProblem}.

Let $\mc{R}$ be the partition (coarse graining) of the phase space $\R^{2\n}$, satisfying $\R^{2\n} = \cup_{R\in\mc{R}} R$, and $R\cap R'=\emptyset$ if $R\neq R'$.
For each region $R$ we define the characteristic function $\chi_R:\R^{2\n}\to\R$,
\begin{equation}
\label{eq:characteristic-function}
\chi_R(\z)=
\begin{cases}
1, \text{ if } \z\in R\\
0, \text{ otherwise}.\\
\end{cases}
\end{equation}

But if we try to contain the wave function in a small region of space, the {\schrod} dynamics will spread it outside that region very fast. For this reason, we will require that the Wigner function is highly peaked inside a coarse-graining region $R$, rather than having its support completely contained in $R$. Hence, instead of the characteristic function of $R$, $\chi_R(\z)$, we will use its convolution product with some highly peaked function $\varphi(\z)$ centered at $(\0,\0)$ in the phase space. Due to the relation with quantum blobs, the natural choice is $\varphi(\z)=\braket{\z}{\0,\0}$, where $\ket{\0,\0}$ is the \emph{coherent state} located at $(\0,\0)$, which is a Gaussian function. We will use the convolution
\begin{equation}
\label{eq:characteristic-convoluted-with-gaussian}
\fqproj{R}=\chi_R\ast\varphi.
\end{equation}

Omn{\`e}s introduced the operators $\cqproj{R}$ corresponding to the Weyl symbols $\fqproj{R}$, and proved that they form a set of quasiprojectors \cite{Omnes1997QuantumClassicalCorrespondenceUsingProjectionOperators}. He gave an equivalent definition,
\begin{equation}
\label{eq:classicality-quasiprojectors}
\cqproj{R} := \frac{1}{\h^\n}\int_R \ket{\x,\p}\bra{\x,\p}\de\x\de\p,
\end{equation}
where $\ket{\x,\p}$ is the \emph{coherent state} centered in the phase space at $(\x,\p)$, defined as a normalized Gaussian Wigner function whose average is the point $(\x,\p)\in\R^{2\n}$.

No two distinct coherent states are orthogonal, so they cannot form a basis of the Hilbert space $\hilbert$, but they form an overcomplete system, providing a \emph{resolution of the identity operator},
\begin{equation}
\label{eq:coherent-resolution}
\wh{I} = \frac{1}{\h^\n}\int_{\R^{2\n}} \ket{\x,\p}\bra{\x,\p}\de\x\de\p = \sum_{R\in\mc{R}}\cqproj{R}.
\end{equation}

Since the characteristic functions $\chi_R$ for all regions in $\mc{R}$ add up to $1$ identically on the phase space, the functions $\fqproj{R}$ from Eq. \eqref{eq:characteristic-convoluted-with-gaussian} form a \emph{partition of unity}, and the corresponding operators $\cqproj{R}$ form, because of linearity, a resolution of the identity operator $\wh{I}$.

Despite the fact that two distinct coherent states always overlap, the operators \eqref{eq:classicality-quasiprojectors} are quasiprojectors, \ie they behave as projectors in a very good approximation, $\cqproj{R_\alpha}\cqproj{R_\beta}\approx \delta_{\alpha\beta}\cqproj{R_\alpha}$, due to the fact that the coarse-graining regions $R$ are large enough \cite{Omnes1997QuantumClassicalCorrespondenceUsingProjectionOperators}. The approximations are of the order of $\hbar^{\sfrac12}$. 
Moreover, the quantum time evolution $e^{-\frac{i}{\hbar}\wh{H}t}\cqproj{R} e^{\frac{i}{\hbar}\wh{H}t}$ of a quasiprojector $\cqproj{R}$ also approximates well the quasiprojector corresponding to the classical evolution of the region $R$ \cite{Omnes1997QuantumClassicalCorrespondenceUsingProjectionOperators}, due to a result by Hagedorn \cite{Hagedorn1980ATimeDependentBornOppenheimerApproximation}.

Omn{\`e}s also showed that there exists a complete sets of actual projectors (\ie idempotent, orthogonal, and adding-up to the identity operator $\wh{I}$)
\begin{equation}
\label{eq:classicality-projectors}
\cproj{R} \approx \cqproj{R}
\end{equation}
that are close to the $\cqproj{R}$ within a similar approximation.

The operators $\(\cqproj{R_\alpha}\)_{R_\alpha\in\mc{R}}$ form the POVM required by the Postulate \ref{ppSQMMacroObjectivity} from Sec. \sref{s:justification}.

Let us recall some basic notions of POVM \cite{NielsenChuang2010QuantumComputationAndQuantumInformation}. If the POVM is defined by the positive semidefinite operators $\(\wh{\Pi}_\alpha\)_\alpha$, $\sum_\alpha\wh{\Pi}_\alpha=\wh{I}$, for each of the operators $\wh{\Pi}_\alpha$ there is an operator $\wh{\Pi}_\alpha^{\sfrac12}$ so that $\wh{\Pi}_\alpha=\wh{\Pi}_\alpha^{\sfrac12\dagger}\wh{\Pi}_\alpha^{\sfrac12}$.
If $\wh{\rho}$ was the state of the measured system before the measurement, after obtaining the outcome $\alpha$ it becomes
\begin{equation}
\label{eq:POVM-outcome}
\wh{\rho}\mapsto \frac{\wh{\Pi}_\alpha^{\sfrac12}\wh{\rho}\wh{\Pi}_\alpha^{\sfrac12\dagger}}{p_\alpha}
\end{equation}
with a probability
\begin{equation}
\label{eq:POVM-outcome-prob}
p_\alpha=\tr\(\wh{\Pi}_\alpha\wh{\rho}_{\ket{\psi}}\).
\end{equation}

In particular, if $\wh{\rho}=\ket{\psi}\bra{\psi}$, the post-measurement state corresponding to the outcome $\alpha$ is $\wh{\Pi}_\alpha^{\sfrac12}\ket{\psi}$. If $\wh{\Pi}_\alpha$ are projectors, then $\wh{\Pi}_\alpha^{\sfrac12}=\wh{\Pi}_\alpha$, because $\wh{\Pi}_\alpha^\dagger=\wh{\Pi}_\alpha$ and $\wh{\Pi}_\alpha=\wh{\Pi}_\alpha^\dagger\wh{\Pi}_\alpha$, and we recover the standard PVM.

In our case, $\cqproj{R}$ are not projectors, but they are positive semidefinite and form a POVM. But there is no measurement as in SQM, each operator $\cqproj{R}$ is used to define the quasirestriction of a state $\ket{\psi}$ to a region $R\in\mc{R}$. We will see later how this yields the usual measurements as in SQM.

\begin{definition}
\label{def:classicality-operators}
We call the operators $\cqproj{R}$ from Eq. \eqref{eq:classicality-projectors} \emph{classicality quasiprojectors}, and the operators $\cproj{R}$ from Eq. \eqref{eq:classicality-projectors} \emph{classicality projectors}.
We say that a Wigner function $W_{\psi}$ is \emph{quasirestricted} to the coarse-graining region $R$ if $\ket{\psi}=\cqproj R^{\sfrac12}\ket{\psi'}$ for some $\ket{\psi'}\in\hilbert$.
\end{definition}
In other words, $\ket{\psi}$ is in the image of $\cqproj R^{\sfrac12}$, $\ket{\psi}\in\im\cqproj R^{\sfrac12}=\cqproj R^{\sfrac12}\hilbert$.
This suggests the following criterion
\begin{criterion}
\label{criterion-classicality}
A quantum state $\ket{\psi}$ is quasiclassical if there is a coarse-graining region $R\in\mc{R}$ so that the Wigner function $W_{\psi}$ is \emph{quasirestricted} to $R$.
\end{criterion}

We can therefore postulate
\setpostulatePStag{PS6}
\begin{postulatePS}[Macro objectivity]
\label{ppPSMacroObjectivity}
For any time, there is a coarse-graining region of the classical phase space within which the quantum state of the total system is quasirestricted.
\end{postulatePS}

\subsection{Dynamical law}
\label{s:postulate-dynamics}

The time evolution of the Wigner function $W_\psi$, corresponding to the {\schrod} evolution \eqref{eq:schrod}, is given by the \emph{{\vonneum} equation}
\begin{equation}
\label{eq:Liouville-von-Neumann}
\frac{\partial{W_\psi(\z,t)}}{\partial t} = -\moyalBracket{W_\psi(\z,t),H(\z,t)},
\end{equation}
where $H(\z,t)$ is the Weyl symbol of the Hamiltonian operator $\wh{H}(t)$ and
\begin{equation}
\label{eq:Moyal-bracket}
\moyalBracket{A,B} :=\frac{1}{i\hbar}\(A\star B - B\star A\)
\end{equation}
 is the \emph{(Groenewold-)Moyal bracket}.

Postulate \ref{ppPSMacroObjectivity} introduced classicality directly, rather than attempting to derive it. Whatever experiment the observer performs, the outcome can be observed only when it produces a macroscopic difference, \ie when the state of the system moves from a coarse-graining region of the phase space to another one. As long as there is no macroscopically observable difference, there should be no observable projection. This justifies
\setpostulatePStag{PS4}
\begin{postulatePS}[Dynamics]
\label{ppPSQuantumDynamics}
The evolution of the quantum state of the system within the same coarse-graining region is given by the {\vonneum} equation.
\end{postulatePS}

\subsection{Quantum transitions}
\label{s:postulate-transitions}

Suppose that at some time $t_0$ the Wigner function $W_{\psi}(\x,\p,t_0)$ is quasirestricted to the coarse-graining region $R_0\in\mc{R}$, and at a future time $t_1>t_0$ it enters, by its evolution according to equation \eqref{eq:Liouville-von-Neumann}, in the regions $R_1,\ldots,R_N\in\mc{R}$. 
On the Hilbert space, this corresponds to the unitary evolution of $\ket{\psi_0}$ until, at $t_1$, it becomes $\ket{\psi_1}=\wh{U}\(t_1,t_0\)\ket{\psi_0}$, which is a linear combination of states quasirestricted to the regions $R_1,\ldots,R_N\in\mc{R}$. In other words,
\begin{equation}
\label{eq:unitary_spread}
\ket{\psi_1}=\sum_{j=1}^kc_j\cqproj{R_j}^{\sfrac12}\ket{\psi_1}\approx\sum_{j=1}^kc_j\cproj{R_j}\ket{\psi_1},
\end{equation}
so the coefficients $c_j$ are approximately
\begin{equation}
\label{eq:unitary_spread_coeff}
c_j\approx\dsfrac{\bra{\psi_1}\cproj{R_j}\ket{\psi_1}}{\braket{\psi_1}{\psi_1}}.
\end{equation}

As Postulate \ref{ppPSMacroObjectivity} specifies, at $t_1$ the Wigner function describing the system will be in only one of the regions $R_1,\ldots,R_N\in\mc{R}$, say $R_j$, $j\in\{1,\ldots,N\}$. We now introduce the Born rule, stating that the probability for the system to end out in the region $R_j$ is
\begin{equation}
\label{eq:QuantumTransition}
p_j=\tr\(\cqproj{R_j}\wh{\rho}_{\ket{\psi_1}}\) = \int_{\R^{2\n}} \fqproj{R}(\z) W_{\psi_1}(\z)\de\z.
\end{equation}

\setpostulatePStag{PS5}
\begin{postulatePS}[Transition]
\label{ppPSQuantumTransition}
When the Wigner function of the total system propagates from one coarse-graining region to others, it transitions to only one of these regions, with a probability given by Eq. \eqref{eq:QuantumTransition}.
\end{postulatePS}

In Sec. \sref{s:projection_postulate} we will see that the standard Projection Postulate is a consequence of these postulates.

Due to the blurry boundaries of the quasiprojectors \eqref{eq:classicality-quasiprojectors}, the separation between dynamics (Postulate \ref{ppPSQuantumDynamics}) and transitions (Postulate \ref{ppPSQuantumTransition}) is a bit complex. There is an interplay between the two, which will be discussed in Sec. \sref{s:interplay}. In fact, the same holds in the SQM, even though apparently its usual formulations strictly separate dynamics and projections.

\subsection{Back to the Hilbert space formulation}
\label{s:hilbert-space}

Postulates \ref{ppPSQuantumState}--\ref{ppPSMacroObjectivity} can be expressed in terms of the more popular Hilbert space formulation.
In fact, the Postulates \ref{ppPSQuantumState}--\ref{ppPSQuantumOperator} were obtained using the Wigner-Weyl correspondence directly from the Postulates \ref{ppSQMQuantumState}--\ref{ppSQMQuantumOperator}.

The remaining postulates rely on the quasiclassicality projectors $\cqproj{R}$, which play the role of the POVM postulated in Postulate \ref{ppSQMMacroObjectivity}. More about this in Sec. \sref{s:projection_postulate}.

\setpostulateHStag{\ref*{ppSQMMacroObjectivity}$'$}
\begin{postulateHS}[Macro objectivity]
\label{ppHSMacroObjectivity}
There is an overcomplete set of quasiclassicality projectors $\{\cqproj{R}|R\in\mc{R}\}$, and for any time, there is $R\in\mc{R}$, such that the quantum state of the system is in the image of $\cqproj{R}^{\sfrac12}$.
\end{postulateHS}

The dynamics is again directly given by the Wigner-Weyl correspondence between the two formulations:
\setpostulateHStag{\ref*{ppSQMQuantumDynamics}$'$}
\begin{postulateHS}[Dynamics]
\label{ppHSQuantumDynamics}
The evolution of the quantum state is given by the {\schrod} equation, as long as it remains in the image of the same quasiclassicality projector.
\end{postulateHS}

\setpostulateHStag{\ref*{ppSQMQuantumTransition}$'$}
\begin{postulateHS}[Transition]
\label{ppHSQuantumTransition}
When the quantum state $\ket{\psi}$ of the system evolves to leave $\im\cqproj R^{\sfrac12}$, becoming a superposition of states in the images of more quasiclassicality projectors $\ket{\psi}=\sum_{j=1}^kc_j\cqproj{R_j}^{\sfrac12}\ket{\psi_1}$, it transitions into a state $\cqproj{R_j}^{\sfrac12}\ket{\psi_1}$ corresponding to only one of the quasiclassicality projectors, with a probability given by Eq. \eqref{eq:QuantumTransition}.
\end{postulateHS}

In Sec. \sref{s:projection_postulate} I will explain in more detail how the Standard Projection Postulate \ref{ppSQMQuantumTransition} is recovered from Postulate \ref{ppPSQuantumTransition}.

%
%
%
%
%
%

\section{Recovering the standard Projection Postulate}
\label{s:projection_postulate}

Since the Projection Postulate is among the fundamental postulates of SQM, this makes the theory rely on measurements, and implicitly on the existence of a quasiclassical system -- the measuring apparatus, or on the existence of an observer.
Some of the founders of QM even thought that the measuring device is projected into a quasiclassical state by the very observer conducting the measurement. This made researchers like Heisenberg \cite{Heisenberg1958PhysicsAndPhilosophy}, von Neumann \cite{vonNeumann1955MathFoundationsQM}, Wigner \cite{Wigner1967RemarksOnTheMindBodyProblemWignersFriend}, Stapp \cite{Stapp2004QuantumTheoryOfMindBrainInterface,Stapp2015QuantumMechanicsMindBrainConnection}, and others \cite{SEP-qt-consciousness} think that consciousness plays a fundamental role in QM.

By contrast, OSQM does not need to appeal to measurements in its fundamental postulates.
Now we need to show that indeed the theory obtained from these postulates is the same as SQM. In \sref{s:hilbert-space} we have seen that most of the postulates of SQM are equivalent to postulates from OSQM. It remains to show that we can derive the Projection Postulate \ref{ppSQMQuantumTransition} from the Postulate \ref{ppPSQuantumTransition}, and recover the predictions of SQM for the process of quantum measurement.

To show this, let us go back to the system composed of the observed system $\mc{S}$ and the measuring apparatus $\mc{M}$ from \sref{s:postulatesSQM}. Let their phase spaces be $\mc{P}_{\mc{S}}$ and respectively $\mc{P}_{\mc{M}}$. The Hilbert space of total system is $\hilbert_{\mc{M}}\otimes\hilbert_{\mc{S}}$, and the corresponding classical phase space is $\mc{P}_{\mc{M}}\oplus\mc{P}_{\mc{S}}$.

A central remark is that the observed system $\mc{S}$ is not directly observed: whatever we learn about its state by measurement, comes in the form of a change of the macro state of the measuring device $\mc{M}$. Not only the measuring device, but the total system is in a quasiclassical state before the measurement, and it ends out in a quasiclassical state after the measurement, due to the Postulate \ref{ppPSMacroObjectivity}. Whatever can be said about the observed system is inferred from the classical states of the total system before and after the measurement. On the total phase space $\mc{P}_{\mc{M}}\oplus\mc{P}_{\mc{S}}$, the following coarse-graining regions associated to the systems $\mc{M}$ and $\mc{S}$ are relevant:
\begin{enumerate}
	\item 
	A coarse-graining region $R_{\text{ready}}$, corresponding to the system $\mc{S}$ being prepared to be measured, and the system $\mc{M}$ being prepared to measure it.
	\item 
	A number of $N$ coarse-graining regions $R_{\text{outcome}=\lambda_j}$, corresponding to the system $\mc{S}$ being in the eigenstate $\ket{j}$, and the measuring apparatus $\mc{M}$ indicating that the outcome of the measurement is $\lambda_j$, for $j\in\{1,\ldots,N\}$.
\end{enumerate}

In addition, we know that the measuring device is, by construction, such that the only way the composed system can evolve by {\schrod} dynamics is in a superposition of the form from Eq. \eqref{eq:measurement_superposition}.

\begin{equation}
\tag{\ref{eq:measurement_superposition}}
\ket{\text{ready}}_{\mc{M}}\otimes\ket{\psi}\mapsto\sum_{j=1}^N c_j \ket{\text{outcome}=\lambda_j}_{\mc{M}}\otimes\ket{j}.
\end{equation}

In OSQM this translates into the fact that the Wigner function of the total system evolved from the coarse-graining region $R_{\text{ready}}$ to the union of the coarse-graining regions $R_{\text{outcome}=\lambda_j}$. By Postulate \ref{ppPSQuantumTransition}, the total system has to transition to only one of the coarse-graining regions $R_{\text{outcome}=\lambda_j}$, with the probability given by Eq. \eqref{eq:QuantumTransition}, which is $\approx\abs{c_j}^2$, with $c_j$ from Eq. \eqref{eq:measurement_superposition}. Hence, the standard Projection Postulate follows as a consequence of the postulates proposed here. 

For example, consider the detection of light by a photographic plate. In this case, the coarse-graining regions $R_{\text{outcome}=\lambda_j}$ correspond to different distinguishable places where the photon is absorbed by the photographic plate. But the source of light may very well be a distant star. Therefore, in all the time when the photon traveled from that star to the plate, the total system evolved through many different coarse-graining regions of the phase space. Then, how is it that the photon coming from that star was not affected, according to Postulate \ref{ppPSQuantumTransition}, by the fact that the entire universe moved through numerous coarse-graining regions? The answer is simple, each coarse-graining region of the entire universe is a Cartesian product of coarse-graining regions of the subsystems, due to Postulate \ref{ppPSComposite}. If the photon is a separate subsystem and does not interact with other systems between the moment of its emission and the moment of its detection, even if many other systems in the universe cross such boundaries and undergo quantum transitions, it is not affected by their part of quasiprojections. So, for our photon, we can safely ignore the history of transitions of the rest of the world from one coarse-graining region to another. 

If the photon is entangled with other particles, the situation has to be analyzed considering the entangled system, as in the EPR experiment \cite{EPR35,Bohm1951TheParadoxOfEinsteinRosenAndPodolsky}.
The coarse-graining region of the entangled system is a region in the Cartesian product of the phase spaces of each of the entangled particles. If none of them undergoes a transition, the photon does not transition either during its travel. In the case one of them transitions, the projection of the entangled system is such that it leaves unchanged the marginal Wigner function corresponding to the photon's phase space. This treatment of a particle from an entangled system is exactly as in the SQM. The only difference is that whatever projection happens in SQM, in OSQM it is due to Postulate \ref{ppPSQuantumTransition}.

\section{The interplay between dynamics and transitions}
\label{s:interplay}

In this section I will explain the interplay between dynamics and transitions. Already in SQM the dynamics is more complex than the standard narrative of unitary evolution interrupted once in a while by projections, as we have seen in Sec. \sref{s:justification}.

But first, let us apply the discussions from Sec. \sref{s:postulate-transitions} and \sref{s:projection_postulate} to the situation when the observed system is an excited atom, assuming that the measuring apparatus is a detector which either detected the decay or not. Two coarse-graining regions are relevant here, call them $R_{\text{excited}}$ and $R_{\text{decayed}}$. As the system is contained in region $R_{\text{excited}}$, the measuring device registers no decay. But as time goes on, the Wigner function spreads out of region $R_{\text{excited}}$, leaking into region $R_{\text{decayed}}$. The \emph{quantum Zeno effect}  \cite{MisraAndSudarshan1977QuantumZenoEffect} implies that the monitoring of the excited atom while the system's Wigner function only spreads very little into region $R_{\text{decayed}}$ projects it back to the excited state, preventing or delaying the decay \cite{Khalfin1958DecayOfAQuasiStationaryStateZenoEffect,WilkinsonEtAl1997NonExponentialDecayInQuantumTunnelingZenoEffect}. By contrast, monitoring it after its Wigner function spread enough into region $R_{\text{decayed}}$ results in an enhancement of the decay rate \cite{Chaudhry2016ZenoAntiZenoEffect}. This exemplifies the interplay between dynamics and transitions, which will be explained now.

The quasiprojectors $\cqproj{R}$ used to define quasiclassicality project the Wigner state only approximately to the coarse-graining regions. Due to the use of the convolutions $\fqproj{R}=\chi_R\ast\varphi$ with the Gaussian function $\varphi$, instead of the characteristic functions $\chi_R$ of the coarse-graining regions, the projection is distorted around the boundaries of the regions.
This means that the separation between dynamics (Postulate \ref{ppPSQuantumDynamics}) and transitions (Postulate \ref{ppPSQuantumTransition}) is not exact. The Wigner functions have ``tails'', of very low values, that go outside of the coarse-graining region $R$. This means that even for the times when the Wigner function of the system is included in a coarse-graining region $R$, quasiprojections by $\cqproj{R}$ happen, albeit with a very small effect. 

Here is a more detailed explanation. For each coarse-graining region $R$, there is an internal region $R^\circ\subset R$, defined by $R^\circ=\fqproj{R}^{-1}(1)$, \ie $R^\circ$ is the set of all $\z\in R$ for which $\fqproj{R}(\z)=1$. Due to the fact that $R$ is much larger than the width of the Gaussian function $\varphi$, the region $R^\circ$ is approximately the same as $R$. Inside $R^\circ$, the Wigner function $W_{\psi}$ satisfies the {\vonneum} equation \eqref{eq:Liouville-von-Neumann}, independently on the fact that $W_{\psi}$ does not vanish completely outside $R^\circ$. The reason is that the Hamiltonian function $H(\x,\p)$ in Eq. \eqref{eq:Liouville-von-Neumann} acts on the Wigner function $W_{\psi}$ through the Moyal product \eqref{eq:Moyal-prod-local}, which is local on the phase space. This means that the dynamics is not affected by the quasiprojection $\cqproj{R}$ for $\z\in R^\circ$, and Postulate \ref{ppPSQuantumDynamics} holds exactly for these points.
But for $\z\notin R^\circ$ near the boundary of $R$, the dynamics is distorted by the projection.

The Wigner function may have a very small value outside the region $R$, because Postulate \ref{ppPSMacroObjectivity} and the classicality quasiprojector $\cqproj{R}$ allow this. Even when $W_{\psi}$ is restricted to the region $R$, its tiny ``tails'' slightly spread outside of $R^\circ$ and then outside of $R$, and activate Postulate \ref{ppPSQuantumTransition} (our version of the Projection Postulate). But since the value of $W_{\psi}$ is very small outside of $R$, the probability to project the state on another region $R'\neq R$ is small, and the Born rule implies that the chosen quasiprojector $\cqproj{R}$ is significantly more often the preferred one, and it projects the Wigner function $W_{\psi}$ back into region $R$. Note that even if the Wigner function is quasiconstrained to the same region $R$ for a certain amount of time, the quantum Zeno effect does not imply that it remains unchanged. The reason why the Wigner function evolves even when it is quasirestricted to $R$ is that the operator $\cqproj{R}$ has a very high degeneracy in the eigenvalue $\lambda=1$. So the quantum Zeno effect does not apply for $\z\in R^\circ$, and Postulate \ref{ppPSQuantumDynamics} indeed holds exactly there.

However, $W_\psi$ continues to evolve towards the boundary of region $R$. As it accumulates at the boundary of region $R$, where the function $\fqproj{R}$ overlaps with the functions $\fqproj{R_j}$, $j\in\{1,\ldots,N\}$ from Sec. \sref{s:postulate-transitions}, it becomes more probable that Postulate \ref{ppPSQuantumTransition} allows $W_\psi$ to transition to another coarse-graining region $R_j$.

This may seem different from the usual formulations of SQM, where the common understanding is that only when quantum measurements happen, the Projection Postulate applies \cite{vonNeumann1955MathFoundationsQM,Dirac1958ThePrinciplesOfQuantumMechanics}. But in reality this alone does not explain the fact that the macro level is quasiclassical, in particular it does not justify the existence of a quasiclassical measuring apparatus to perform the postulated measurement in the first place. The Copenhagen Interpretation, the default companion of the SQM, explains this by the presence of the observer, whose sensory organs (or consciousness?) act like measuring devices. The observer is the one who makes the measuring device be quasiclassical, and the one who, by monitoring the measuring device, maintains it to be quasiclassical rather than to evolve into a superposition. So even SQM has this interplay between the dynamics and the projection. In OSQM this interplay is visible, and even for the case when the system is expected to simply follow the evolution equation, this happens strictly only inside region $R^\circ\subset R$, and for other parts of the phase space the evolution involves a continuous projection.

In conclusion, a main difference between the Postulates HS and the Postulates PS is that the latter lead to gradual quantum jumps, rather than instantaneous, discontinuous ones. This difference leads to different experimental predictions of how the projection takes place, and will be discussed in Sec. \sref{s:experiment}.

\section{Experimental predictions}
\label{s:experiment}

At first sight, it would seem that OSQM does not lead to significantly different predictions compared to SQM. The discussion from Sec. \sref{s:justification}, particularly Postulate \ref{ppSQMMacroObjectivity}, suggests that only reasonable constraints were imposed to SQM, constraints that are already present in practice, due to the Objectivity Problems \ref{pbo:zeno}--\ref{pbo:macrolevel}. But in fact the interplay between Postulates \ref{ppPSQuantumDynamics}, \ref{ppPSQuantumTransition}, and \ref{ppPSMacroObjectivity}, described in Sec. \sref{s:interplay}, leads to the conclusion that the wave function happens gradually, rather that all at once. Since the propagation of the Wigner function from a coarse-graining region to the others is continuous, and since the regions are characterized by quasiprojectors that have smooth Weyl symbols on the phase space, Postulate \ref{ppPSQuantumTransition} implies that the projections happen in small rates, almost continuously. This is in stark contrast with the usual formulations of SQM, where the projections is described as taking place suddenly, discontinuously.

Fortunately, recent experiments already verified the expectation that quantum jumps happen discontinuously, and they showed that it actually happens in a continuous and controllable manner \cite{MinevEtAl2019ToCatchAndReverseAQuantumJumpMidFlight,Minev2019ToCatchAndReverseAQuantumJumpMidFlight}. The team could monitor the jumps, could anticipate them, and even reverse them. 
The result of the experiment does not contradict the SQM Postulates \ref{ppSQMQuantumState}--\ref{ppSQMQuantumTransition}, only the spread expectation that the transitions are sudden and discontinuous (it worth mentioning that {\schrod} objected the existence of discontinuous jumps \cite{Schrodinger1952AreThereQuantumJumpsPartI,Schrodinger1952AreThereQuantumJumpsPartII}, and he thought they are continuous and coherent, just as the experiment showed).
These results are as predicted by the interplay between Postulates \ref{ppPSQuantumDynamics}, \ref{ppPSQuantumTransition}, and \ref{ppPSMacroObjectivity} described in Sec. \sref{s:interplay}, that the transitions happen gradually.

Another prediction, different from SQM, comes straight from Postulate \ref{ppPSMacroObjectivity}, even in its weaker form as Postulate \ref{ppSQMMacroObjectivity}: there are no macro level {\schrod} cats, because all simultaneous measurements are compatible, hence objective.
The reason is that, according to Postulate \ref{ppPSMacroObjectivity}, the Wigner function of the total system is at any times restricted to a coarse-graining region of the classical phase space. In other words, quantum macro states are indistinguishable from classical macro states. Since classically a cat cannot simultaneously be in two distinct macro states, Postulate \ref{ppPSMacroObjectivity} implies that the same is true for the cats in a quantum world. A cat can be in a superposition of different quantum states, of course, but only as long as this superposition can be expressed as consisting of quantum states that are macroscopically indistinguishable. This eliminates the possibility of {\schrod} cats.

Wigner's thought experiment is similar to the {\schrod} cat experiment, except that the cat is replaced by a human who makes a quantum measurement. Since the possible outcomes lead to distinct macro states of the measuring device, a human reading the outcome of the experiment will also be in one of two or more distinct macro states. 
Wigner is supposed to know that his friend is measuring the spin $\sigma_z$ of a particle whose initial state is $\dsfrac{1}{\sqrt 2}\(\ket{0}_z + \ket{1}_z\)$.
By Wigner's knowledge, the result is a superposition, hence the measuring device and Wigner's friend are in a superposition as well.
But this is Wigner's subjective knowledge. But Postulate \ref{ppPSMacroObjectivity} implies that there is a definite macro state, hence a definite outcome is already obtained, and this is an objective fact.
Postulate \ref{ppPSMacroObjectivity} prevents Wigner's friend to be in a superposition of such states, just like in the case of the cat, even if Wigner or anyone else does not look inside the laboratory.

It may seem implausible to find out flesh-and-bones Wigner friend states even if they would be possible in principle, but nevertheless such thought experiments and even physically realized experiments are discussed, and no-go theorems are deduced from them \cite{FrauchigerRenner2018QMNoConsistentDescription,Brukner2018NoGoObserverIndependentWignerFriend,BongEtAl2020StrongNoGoTheoremWignerFriend}.

It must be mentioned that, while this solution avoids {\schrod} cats and Wigner's friend paradox, it may seem to contradict the Legget-Garg theorem \cite{LeggettGarg1985QMvsMacrorealism,Leggett2008Realism,EmaryNeillFranco2013LeggettGarg}. According to this theorem, it is not always possible to simultaneously satisfy the following conditions:
\begin{enumerate}
	\item 
	\emph{Macrorealism}: Macroscopic objects are at all times in only one definite macro state.
	\item 
	\emph{Noninvasive measurement}: It is possible to determine the macro state of the system without disturbing it.
\end{enumerate}
Since Postulate \ref{ppPSMacroObjectivity} is in fact the condition of macrorealism, it follows that what we give up here in case of tension between the two is the second condition, that all measurements are noninvasive. But there is no known situation in which this would affect the possibility of noninvasive observation of the macro states of cats or humans like Wigner's friend.

Wigner's friend paradox is a problem for interpretations that give the observer a central role. Other approaches, like collapse theories \cite{GhirardiRiminiWeber1986GRWInterpretation,sep-qm-collapsetheories}, or the de Broglie-Bohm theory \cite{Bohm52,Bohm2004CausalityChanceModernPhysics}, can be argued to be ``without observers'' \cite{Goldstein1998QuantumTheoryWithoutObserversI,Goldstein1998QuantumTheoryWithoutObserversII}.
However, the paradox is considered to be relevant in general, since it challenges the notion of objective reality in all theories. While some approaches, including OSQM, avoid it, two possible loopholes of the solution proposed here to Wigner's friend should not be ignored though.
One of them is the possibility that Wigner's friend is a microorganism, small enough so that its states are not distinguishable at the macro level. Recent results indicate that a bacterium can be placed in superposition \cite{Marletto2018EntanglementBacteria}, but there is no evidence yet that a bacterium is able to conduct quantum measurements like Wigner's friend is supposed to. The second possibility, assuming that the Strong AI Thesis is true, is that Wigner's friend is a sentient Artificial Intelligence. Then, if quantum computing is scalable enough to support superpositions of AI agents, Wigner's friend paradox can be resurrected. In this case, Postulate \ref{ppPSMacroObjectivity} would imply an objective world, which is the macro quasiclassical one, but also subjective facts for the AI agents.
However, the same argument can be brought in a classical world. In a classical world it is possible to simulate a quantum computer to any desired degree of approximation. Therefore, if Strong AI holds, even a classical computer can support superpositions of Wigner's friend. This renders irrelevant the whole argument that AI Wigner's friend challenges objectivity, because it does not make a quantum world less objective than a classical world.

\section{Discussion}
\label{s:discussion}

In this section I discuss what is achieved by OSQM, and what are some open problems, and possible criticism or objections.

\emph{Problem \pbref{pb:Observer}.}
OSQM does not use the Projection Postulate \ref{ppSQMQuantumTransition} as fundamental, replacing it with Postulate \ref{ppPSQuantumTransition}, which makes no reference to measurements. Nevertheless, we have seen in Sec. \sref{s:projection_postulate} that the Projection Postulate \ref{ppSQMQuantumTransition} can be recovered.

More precisely, in Sec. \sref{s:projection_postulate} we modeled measuring devices as quantum systems in definite quasiclassical macro states, and measurements as physical processes obeying Postulates \ref{ppPSQuantumState}--\ref{ppPSMacroObjectivity}, rather than being fundamental ingredients of the postulates.
A measuring device is just a physical system whose macro states become correlated, by the application of Postulate \ref{ppPSQuantumDynamics}, with the states of the observed system, and then only one of the possible definite macro states is obtained, due to Postulates \ref{ppPSMacroObjectivity} and \ref{ppPSQuantumTransition}.
The observers play no role either. This does not mean that observers are excluded from the theory, just that they do not play any fundamental role, as required by the \emph{Copernican principle}. For the sake of describing experimental procedures, the functionality of observers is to discriminate the states of other systems, being them directly observed or measuring devices, and to keep in their memory records of the distinctions. In other words, the relevant functionality of an observer in experiments is identical to that of a measuring apparatus. This is by no means an attempt to reduce observers to measuring devices, but any further refinement of their description is relegated to other sciences, like Biology, Psychology, and maybe Artificial Intelligence.

Omn{\`e}s proposed and studied quasiprojection operators in the context of the \emph{Consistent Histories} approach to QM. The Consistent Histories approach was developed independently by Griffiths \cite{Griffiths1984ConsistentHistories}, Omn{\`e}s \cite{Omnes1988ConsistentHistoriesLogicalReformulationOfQuantumMechanicsIFoundations,Omnes1992ConsistentInterpretationsQuantumMechanics,Omnes1999UnderstandingQuantumMechanics}, and, from a cosmological perspective and under the name \emph{Decoherent Histories}, by Gell-Mann and Hartle \cite{GellMannHartle1990DecoheringHistories,GellMannHartle1990QuantumMechanicsInTheLightOfQuantumCosmology}. It relies on \emph{frameworks}, \ie algebras of projectors or quasiprojectors, which are used to represent events and histories. Multiple frameworks are used, and the choice of a particular framework is dependent on the questions we ask, on the properties that are measured or that are relevant in the description of a particular history. Here \emph{Quantum Logic} becomes relevant, due to the incompatibility of different frameworks \cite{BirkhoffVonNeumann1936TheLogicOfQuantumMechanics,Isham1994QuantumLogicConsistentHistories}. The evolution is stochastic, and the probabilities for different histories are assigned in the same framework according to the Born rule and the {\schrod} equation. Frameworks can be refined or coarser. Quasiclassical histories are considered to emerge due to decoherence.
OSQM also uses quasiprojectors, as defined by Omn{\`e}s. They correspond to the classical macro states. In the Consistent Histories parlance, we can say that OSQM uses a single ``framework'', which, rather than emerging by decoherence, is defined by the classical coarse graining of the phase space into macro states, and it includes a single history.

\emph{Problem \pbref{pb:ClassicalCriterion}.}
Criterion \ref{criterion-classicality} is the proposed criterion of classicality.

\emph{Problem \pbref{pb:ClassicalLevel}.}
This question received here only a provisional answer: ``because Postulate \ref{ppPSMacroObjectivity} requires quantum states to satisfy Criterion \ref{criterion-classicality}''. The task of this article was to remove the measurements from the formulation of SQM. But a real solution to Problem \ref{pb:ClassicalLevel} was not provided here. A fully satisfactory solution requires most likely an extension of SQM, or a so-called ``interpretation''. As explained in Sec. \sref{s:intro}, even if we appeal to decoherence, it should be done in conjunction with an interpretation where the branching of the wave function is present, like Everett's interpretation, Consistent Histories, or the de Broglie-Bohm interpretation. For one of its problems, that of the preferred basis, one may find that Criterion \ref{criterion-classicality} is more appropriate.

\emph{Problem \pbref{pb:ProjectionCircularity}.}
The circularity mentioned in Problem \ref{pb:ProjectionCircularity} is resolved because the Projection Postulate \ref{ppSQMQuantumTransition} is replaced with Postulate \ref{ppPSQuantumTransition}, which no longer is responsible for the macro level, this one making the object of Postulate \ref{ppPSMacroObjectivity}.

\emph{Problem \pbref{pb:ProjectionCriterion}.}
In OSQM, Postulate \ref{ppPSQuantumTransition} solves Problem \ref{pb:ProjectionCriterion}. Following the discussion in Sec. \sref{s:interplay}, we have seen that projection occurs to some extent continuously.

\emph{Problem \pbref{pb:StateReality}.}
The answer to this question is yes. In OSQM, the Wigner function is well defined at all moments of times. While here we are not interested in the problem of ontology, this makes possible (but not compulsory) to assign an ontology to the wave function or to the Wigner function. The wave function is defined on the configuration space, while the Wigner function on the phase space. But if by ``ontology'' we mean something defined on the physical $3$-dimensional space, it is possible to represent the wave function in terms of fields defined on the $3$-dimensional space, and in \cite{Sto2019RepresentationOfWavefunctionOn3D} was given such a representation, which serves at least as a proof of concept. Therefore, since OSQM allows quantum states to be well defined at all times, it also allows an ontology on the $3$-dimensional physical space, as shown by the construction made in \cite{Sto2019RepresentationOfWavefunctionOn3D}.

Even though the standard Projection Postulate was derived from the Postulates presented here, one may object that OSQM is not merely a reformulation of SQM, but an entirely different collapse theory. I think that this depends on what interpretations we adopt for the SQM and of the Copenhagen Interpretation.

\emph{Problem \pbref{pb:WignerFriend}.}
By being slightly more restrictive than SQM, OSQM does not allow macro \emph{{\schrod} cats} or flesh-and-bones \emph{Wigner friends} \cite{Wigner1967RemarksOnTheMindBodyProblemWignersFriend}, due to the Postulate \ref{ppPSMacroObjectivity}. This prevents the potential problems attributed to SQM recently by Frauchiger and Renner \cite{FrauchigerRenner2018QMNoConsistentDescription}, by appealing to thought experiments based on Wigner's friend. The possibility of AI Wigner friends running on a quantum computer \cite{BongEtAl2020StrongNoGoTheoremWignerFriend} is not excluded, but this is not a problem for OSQM, which is not based on observers. If the Strong AI Thesis is possible, and if quantum computing able to support superpositions of such AIs turns out to be consistent with the coarse graining invoked in Postulate \ref{ppPSMacroObjectivity}, then AI-type Wigner's friends may exist, without violating any of the Postulates.
A better understanding of the coarse graining is essential for understanding the limits it imposes to quantum coherence and quantum computing.

\textbf{Acknowledgement.}
The author thanks Eric Cavalcanti, Eliahu Cohen, Maurice de Gosson, Danko Georgiev, Peter W. Morgan, Igor Salom, Larry Schulman,
and the anonymous referees,
for their valuable comments and suggestions offered to a previous version of the manuscript. Nevertheless, the author bears full responsibility for the article.

%

\end{document}